\RequirePackage[2020-02-02]{latexrelease}
\documentclass[aps,prd]{revtex4}

\usepackage{amsmath}
\usepackage{amssymb}
\usepackage{bm}
\usepackage{graphicx}
\usepackage{multirow}

\usepackage{textcomp}

\allowdisplaybreaks[1]

\begin{document}

\title{Dispersive determination of electroweak-scale masses}

\author{Hsiang-nan Li}
\affiliation{Institute of Physics, Academia Sinica,
Taipei, Taiwan 115, Republic of China}

\date{\today}

\begin{abstract}
We demonstrate that the Higgs boson mass can be extracted from the dispersion relation obeyed 
by the correlation function of two $b$-quark scalar currents. The solution to the dispersion 
relation with the input from the perturbative evaluation of the correlation function up to 
next-to-leading order in QCD and with the $b$ quark mass $m_b=4.43$ GeV demands a specific 
Higgs mass 114 GeV. Our observation offers an alternative resolution to the long-standing 
fine-tuning problem of the Standard Model (SM): the Higgs mass is determined dynamically 
for the internal consistency of the SM. The similar formalism, as applied to the correlation 
function of two $b$-quark vector currents with the same $m_b$, leads to the $Z$ boson mass 
90.8 GeV. This solution exists only when the $Z$ and $W$ boson masses are proportionate, 
conforming to the Higgs mechanism of the electroweak symmetry breaking. We then consider the 
mixing between the $Q\bar u$ and $\bar Qu$ states for a fictitious heavy quark $Q$ and a $u$ 
quark through the $b\bar b$ channel, inspired by our earlier analysis of neutral meson mixing. 
Its dispersion relation, given the perturbative input from the responsible box diagrams and 
the same $m_b$, fixes the top quark mass 176 GeV. It is highly nontrivial to predict the 
above electroweak-scale masses with at most 9\% deviation from their measured values using 
the single parameter $m_b$. More accurate results are expected, as more precise perturbative 
inputs are adopted.

\end{abstract}

\maketitle

\section{INTRODUCTION}

Recently we proposed the possibility that the parameters in the Standard Model (SM) are not free, 
but arranged properly to achieve the internal dynamical consistency \cite{Li:2023dqi}. This proposal 
was motivated by the emergence of a heavy (charm or bottom) quark mass in the dispersive analysis 
on decay widths of a heavy meson $H_Q$ formed by a fictitious heavy quark $Q$ with an arbitrary 
mass $m_Q$. A decay width, as the absorptive piece of a heavy meson matrix element 
of the four-quark effective operators, satisfies a dispersion relation, which imposes a stringent 
connection between high-mass and low-mass behaviors of a decay width. A solution to the 
dispersion relation with the input from heavy quark expansion (HQE) at high mass specifies the 
value of $m_Q$, which turns out to coincide with the physical $c$ or $b$ quark mass. Starting 
with massless final-state up and down quarks, we have shown that the solution for 
the decay $Q\to du\bar d$ ($Q\to c\bar ud$) with the leading-order HQE input leads to the 
$c$ ($b$) quark mass $m_c=1.35$ ($m_b= 4.4$) GeV, given the binding energy $\bar\Lambda=0.5$ 
(0.6) GeV of a heavy meson. Requiring that the dispersion relation for the $Q\to su\bar d$ 
($Q\to d\mu^+\nu_\mu$, $Q\to u\tau^-\bar\nu_\tau$) decay yields the same heavy quark mass, we 
obtained the strange quark (muon, $\tau$ lepton) mass $m_s= 0.12$ GeV ($m_\mu=0.11$ GeV, 
$m_\tau= 2.0$ GeV). The above particle masses, close to the measured ones, support the 
aforementioned proposal.
 
This paper will further demonstrate that the electroweak-scale masses, i.e., the Higgs boson, 
$Z$ boson and top quark masses, can also be determined in a similar manner. To extract the Higgs 
boson mass, we investigate the dispersion relation obeyed by the correlation function of two 
$b$-quark scalar currents, like those employed in QCD sum rules \cite{SVZ} for 
probing resonance properties (see \cite{Narison:1994ds}, for instance). The solution to the 
dispersion relation, with the input from the perturbative evaluation of the correlation function 
up to next-to-leading order (NLO) in QCD and with the $b$ quark mass $m_b=4.43$ GeV (slightly 
higher than that from \cite{Li:2023dqi}), demands a specific scalar mass 114 GeV, close to 
the measured one $m_H=(125.25\pm 0.17)$ GeV \cite{PDG}. This observation offers an alternative 
resolution to the long-standing fine-tuning problem of the SM: the Higgs boson must have this 
mass to make the internal consistency of the SM dynamics. The above formalism, as applied to the 
correlation function of two $b$-quark vector currents with the same $m_b$, generates the $Z$ boson 
mass 90.8 GeV, close to the measured one $m_Z=(91.1876\pm 0.0021)$ GeV \cite{PDG}. We then consider 
the mixing between the $Q\bar u$ and $\bar Qu$ states for a fictitious heavy quark $Q$ and a $u$ 
quark through the $b\bar b$ channel, which was inspired by our earlier dispersive analysis on 
neutral meson mixing \cite{Li:2022jxc}. Its dispersion relation, given the perturbative input from 
the responsible box diagrams and the same $m_b$, fixes the top quark mass 176 GeV, close to the 
measured one $m_t=(172.69\pm 0.30)$ GeV \cite{PDG}. The solved particle masses with the 
single parameter $m_b$, deviating from the measured values by at most 9\%, are highly nontrivial. 
Our studies suggest that the masses over a huge hierarchy in the SM, from 0.1 GeV to 100 GeV, 
can be correlated to each other through appropriate dispersion relations.


\section{HIGGS BOSON MASS}

\begin{figure}
\begin{center}
\includegraphics[scale=0.35]{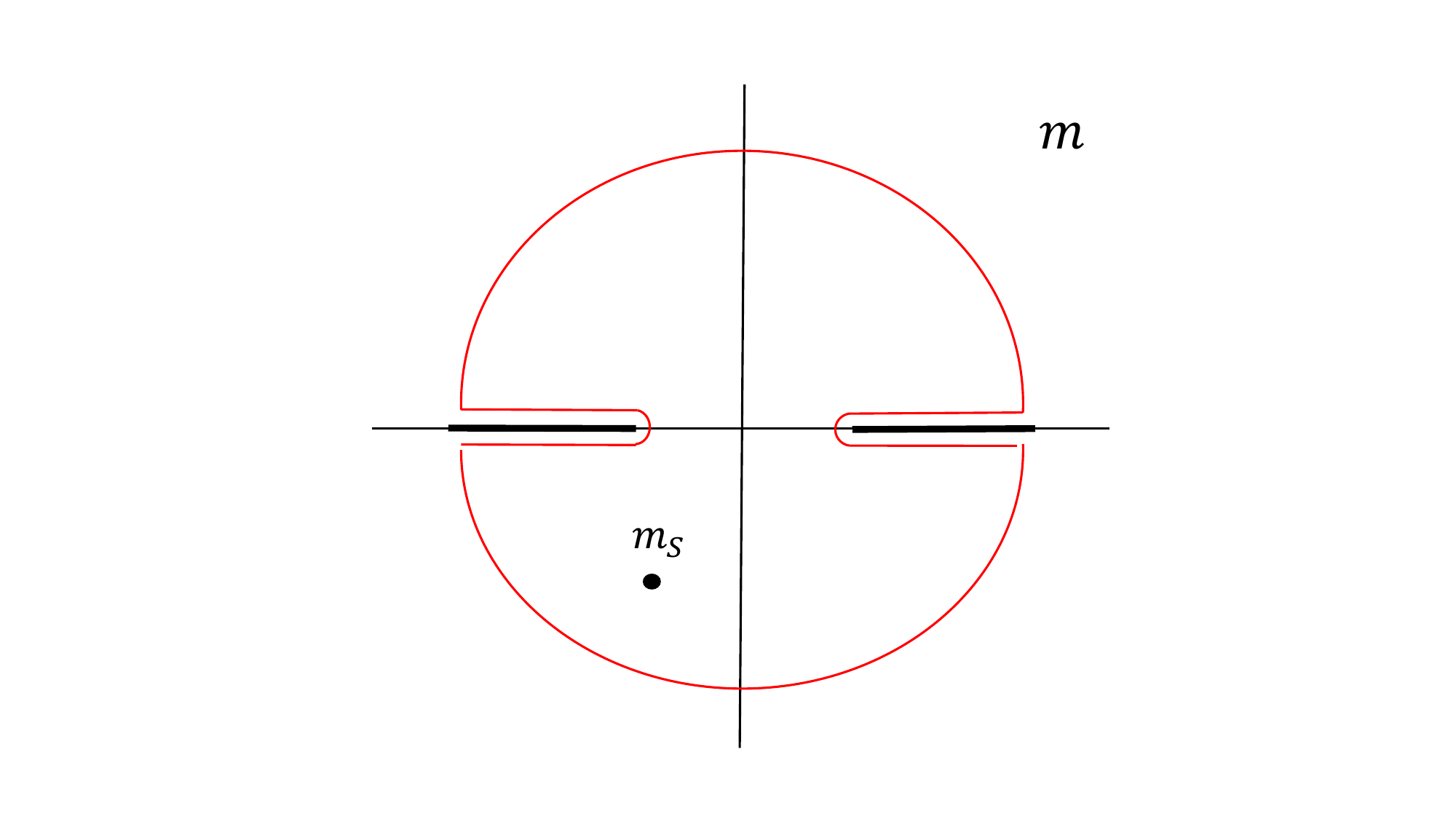}
\caption{\label{fig1}
Contour for Eq.~(\ref{con}), where the thick lines represent the branch cuts.}
\end{center}
\end{figure}

The nonperturbative approach based on dispersion relations for physical observables was proposed 
in \cite{Li:2020xrz}, and applied to the explorations of various quantities in
\cite{Li:2020fiz,Li:2020ejs,Li:2021gsx,Li:2022qul}. Following the threads of the above references,
we construct the dispersion relation obeyed by the two-point correlation function
\begin{eqnarray}
\Pi(q^2)=i\int d^4xe^{iq\cdot x}
\langle 0|T[J(x)J(0)]|0\rangle,\label{cur}
\end{eqnarray}
with the momentum $q$ being injected into the $b$-quark scalar current $J=\bar b b$. The momentum 
squared $q^2$ defines an invariant mass $m_S=\sqrt{q^2}$ of the scalar attaching to the current. 
The standard operator product expansion (OPE) implemented in QCD sum rules is reliable for 
Eq.~(\ref{cur}) in the large $m_S$ region. Inserting various states between the two
operators $J(x)$ and $J(0)$, one realizes that information of any physical particle allowed
to decay into a $b\bar b$ quark pair can be extracted from the above correlation function
in principle, as the OPE is known precisely enough.
It is thus clear that the $b\bar b$ final state, instead of lighter ones like 
$c\bar c$, is appropriate for extracting the Higgs boson mass, because no other
scalar particles with masses above the $B$-meson-pair threshold decay into $b\bar b$.

It has been noticed \cite{Generalis:1983hb,Bagan:1994dy} that the power corrections 
from heavy-quark condensates can be absorbed into those from the gluon condensates. 
The gluon condensate effects, starting with $\langle\alpha_sG^2\rangle\sim O(0.1)$ 
GeV$^4$ \cite{Wang:2016sdt,Narison:2014wqa} and down by powers of $m_S>m_b$, are negligible. 
Hence, we keep only the leading term in the OPE, i.e., the perturbative contribution 
\cite{Gorishnii:1983cu,Mihaila:2015lwa,Kanemura:2019kjg}, which gives rise to the 
spectral function
\begin{eqnarray}
{\rm Im}\Pi^{\rm p}(m_S)\propto \frac{(m_S^2-4m_b^2)^{3/2}}{m_S}
\left[1+\frac{\alpha_s(\mu)}{\pi}C_F
\left(\frac{17}{4}+\frac{3}{2}\ln\frac{\mu^2}{m_S^2}\right)\right].
\label{p1}
\end{eqnarray}
The argument of the above function has been changed to $m_S$, and the suppressed constant 
prefactor is not crucial for the reasoning below. For the NLO correction, we include only 
that in QCD for illustration, which dominates over others from the electroweak interaction. 
Equation~(\ref{p1}) is nothing but the width for a scalar decay into a $b\bar b$ quark pair 
without the initial particle density factor $1/(2m_S)$ and the relevant Yukawa coupling constant.

Viewing the presence of $1/m_S$ in Eq.~(\ref{p1}), we consider the contour integration of 
$m_S^2\Pi(m_S)$ in the complex $m_S$ plane, instead of the $m_S^2$ plane, which contain 
different branching cuts \cite{Li:2023dqi}. This manipulation facilitates the derivation of 
a solution to the dispersion relation as seen later. The contour consists of two 
pieces of horizontal lines above and below the branch cut along the positive real axis, two pieces 
of horizontal lines above and below the branch cut along the negative real axis, and a circle $C_R$ 
of large radius $R$ as depicted in Fig.~\ref{fig1}. We then have the identity
\begin{eqnarray}
\frac{1}{2\pi i}\oint \frac{m^2\Pi(m)}{m-m_S}dm
=m_S^2\Pi^{\rm p}(m_S)=\frac{1}{2\pi i}\oint \frac{m^2\Pi^{\rm p}(m)}{m-m_S}dm.\label{con}
\end{eqnarray}
The residue from the pole at $m=m_S$ marked in Fig.~\ref{fig1} has been 
replaced by its perturbative expression $\Pi^{\rm p}$, since perturbation holds well for high 
scales. This residue is further written as the contour integral of
$m_S^2\Pi^{\rm p}(m_S)$ based on the analyticity of perturbative calculations
\cite{Li:2021gsx,Li:2022jxc}. An intermediate advantage of considering the integrand $m^2\Pi(m)$ is 
apparent: if there exists a singularity of $\Pi(m)$ in the low $m$ region, its 
residue will be suppressed by a power of $m^2/m_S^2$ relative to the one in Eq.~(\ref{con}),
and can be dropped.

Canceling the contributions from the large circle $C_R$ on both sides of Eq.~(\ref{con}), 
which are approximated by the perturbative ones reliably, we arrive at  
\begin{eqnarray}
\int_{2m_B}^R\frac{m^2{\rm Im}\Pi(m)}{m-m_S}dm-
\int^{-2m_B}_{-R}\frac{m^2{\rm Im}\Pi(m)}{m-m_S}dm
=\int_{2m_b}^R\frac{m^2{\rm Im}\Pi^{\rm p}(m)}{m-m_S}dm-
\int^{-2m_b}_{-R}\frac{m^2{\rm Im}\Pi^{\rm p}(m)}{m-m_S}dm.\label{mas}
\end{eqnarray}
The hadronic threshold $2m_B$, $m_B$ being the $B$ 
meson mass, and the quark-level threshold $2m_b$ have been assigned. The imaginary part of the 
perturbative correlation function is attributed to the logarithmic factor $\ln(-m^2+4m_b^2)$
from the $b$ quark loop, which can be split into $\ln(-m^2+4m_b^2)=\ln (m+2m_b) + \ln(-m+2m_b)$. 
The former (latter) gives the phase $i\pi$ ($-i\pi$) for $m<-2m_b$ ($m>2m_b$) along the contour 
above the branch cut. This opposite sign has been reflected explicitly between the two 
terms on each side of Eq.~(\ref{mas}). The unknown spectral function ${\rm Im}\Pi(m)$, 
involving strong dynamics characterized by the scale $m_B$, will be solved from the 
above dispersion relation with the input of ${\rm Im}\Pi^{\rm p}(m)$. That is, Eq.~(\ref{d4}) 
will be handled as an inverse problem \cite{Li:2020xrz}.

Note that the input ${\rm Im}\Pi^{\rm p}(m)$ is an odd function in $m$, so is the corresponding 
solution ${\rm Im}\Pi(m)$ from Eq.~(\ref{mas}). The variable change $m\to -m$ for the 
second integrals on both sides leads to
\begin{eqnarray}
& &\int_{4m_B^2}^{R^2}\frac{m{\rm Im}\Pi(m)}{m_S^2-m^{2}}dm^{2}=
\int_{4m_b^2}^{R^2}\frac{m{\rm Im}\Pi^{\rm p}(m)}{m_S^2-m^{2}}dm^{2}.\label{d4}
\end{eqnarray}
We supply an intuitive interpretation for the above equation. The unknown spectral function
${\rm Im}\Pi(m)$ on the left-hand side collects contributions from real physical particles.
Take the spectral function in the case of light-quark vector currents as an example.
As the momentum squared $q^2$ injected into the vector current increases from zero and crosses 
the di-pion threshold, the two virtual quarks emitted from one current fragment into two real 
pions, whose invariant mass defines the hadronic threshold $2m_\pi$. As $q^2$ increases 
further and crosses the $\rho$ meson threshold, the two virtual quarks emitted from one current 
can "annihilate" into a real $\rho$ meson, which then "decays" into the two virtual quarks 
that end at the other current. One can think of the coupling between the $\rho$ meson and the 
quark pair as $g_{\rho q\bar q}$ appearing in the Nambu-Jona-Lasinio model 
\cite{Polleri:1996rw}. This real $\rho$ meson contribution to the unknown spectral function 
has been parametrized as a pole term in QCD sum rules.

In the present case with the $b$ quark scalar currents, the hadronic threshold $2m_B$ in 
Eq.~(\ref{mas}) corresponds to $2m_\pi$, and a real Higgs boson contribution
corresponds to the real $\rho$ meson contribution mentioned above, with the Yukawa coupling
between the Higgs boson and the $b$-quark pair corresponding to $g_{\rho q\bar q}$.
The right-hand side of Eq.~(\ref{d4}) is evaluated in perturbation theory systematically
by starting with two real $b$ quarks. Then Eq.~(\ref{d4}) means that the two
dispersive integrals, one in terms of the unknown spectral function from the contributions of real 
physical particles and the other in terms of the perturbative spectral function from the 
contribution of real $b$ quarks, should be equal at large $m_S$.  
Solving Eq.~(\ref{d4}) directly, one can extract the Higgs boson mass in the same way 
as extracting the $\rho$ meson mass from the correlation function of light-quark currents 
\cite{Li:2021gsx}. The idea behind our formalism is thus similar to that of QCD sum rules, 
but with power corrections in $(m_B-m_b)/m_S$ originating from the difference between 
the thresholds $2m_B$ and $2m_b$, which are necessary for establishing a physical solution 
\cite{Li:2022jxc}. Without these power corrections, i.e., if $m_b$ is equal to $m_B$, there will 
be only the trivial solution ${\rm Im}\Pi(m)={\rm Im}\Pi^{\rm p}(m)$ and no constraint on the 
Higgs boson mass.

It has been proved rigorously \cite{Xiong:2022uwj} that the solution to an integral equation 
like Eq.~(\ref{d4}) (the Fredholm equation of the first kind), if existing, is unique 
under adequate boundary conditions. We will construct a solution to Eq.~(\ref{d4}) below. 
Moving the integrand on the right-hand side to the left-hand side, and regarding it 
as a subtraction term, we get 
\begin{eqnarray}
& &\int_{4m_b^2}^\infty\frac{\Delta\rho(m)}{(m_S^2-m^{2})}dm^2=0,\nonumber\\
& &\Delta\rho(m)\equiv m\Delta\Pi(m),\;\;\;\;
\Delta\Pi(m)={\rm Im}\Pi(m)-{\rm Im}\Pi^{\rm p}(m).\label{ge}
\end{eqnarray}
The subtracted unknown function $\Delta\Pi(m)$ is fixed to $-{\rm Im}\Pi^{\rm p}(m)$ in the 
interval $(2m_b,2m_B)$ of $m$, and approaches zero at large $m$, where 
${\rm Im}\Pi(m)\to{\rm Im}\Pi^{\rm p}(m)$. This explains why the upper bound $R^2$ of the 
integration variable $m^2$ can be extended to infinity. 
The procedure of solving Eq.~(\ref{ge}) has been elucidated in \cite{Li:2023dqi}, which is  
recaptured here for a self-contained presentation. We change $m_S$ and $m$ in Eq.~(\ref{ge}) into 
the dimensionless variables $u$ and $v$ via $m_S^2-4m_b^2=u\Lambda$ and $m^2-4m_b^2=v\Lambda$, 
respectively, obtaining
\begin{eqnarray}
\int_{0}^\infty dv\frac{\Delta\rho(v)}{u-v}=0.\label{i2}
\end{eqnarray}
The purpose of introducing the arbitrary scale $\Lambda$ will become transparent shortly. 

Because $\Delta\rho(v)$ diminishes at large $v$; namely, the major contribution to Eq.~(\ref{i2}) 
comes from the region with finite $v$, we can expand Eq.~(\ref{i2}) into a power series in 
$1/u$ for sufficiently large $|u|$ by inserting $1/(u-v)=\sum_{i=1}^\infty v^{i-1}/u^i$.
Note that $u$, i.e., $m_S$ could be a complex number, as indicated in Fig.~\ref{fig1}. 
Equation~(\ref{i2}) then dictates a vanishing coefficient for every power of $1/u$. 
We start with the case with $N$ vanishing coefficients,
\begin{eqnarray}
\int_{0}^\infty dvv^{i-1}\Delta\rho(v)=0,\;\;\;\;i=1,2,3\cdots,N,\label{i3}
\end{eqnarray}
where $N$ will be pushed to infinity eventually. The first $N$ polynomials $L_{0}^{(\alpha)}(v)$, 
$L_{1}^{(\alpha)}(v)$, $\cdots$, $L_{N-1}^{(\alpha)}(v)$ are composed of 
the terms $1$, $v$, $\cdots$, $v^{N-1}$ appearing in the above expressions. 
Equation~(\ref{i3}) thus implies an expansion of $\Delta\rho(v)$ in terms of the generalized 
Laguerre polynomials $L_j^{(\alpha)}(v)$ with degrees $j$ not lower than $N$,
\begin{eqnarray}
\Delta \rho(v)=\sum_{j=N}^{N'} a_{j}v^\alpha e^{-v}L_{j}^{(\alpha)}(v),\;\;\;\;
N'>N,\label{d0}
\end{eqnarray}
owing to the orthogonality of the polynomials, in which $a_{j}$ represent a set of unknown 
coefficients. The index 
$\alpha$ describes the behavior of $\Delta \rho(v)$ around $v\sim 0$. The highest degree $N'$ 
can be fixed in principle by the aforementioned initial condition of $\Delta \rho(v)$ in the interval 
$(0,4(m_B^2-m_b^2)/\Lambda)$ of $v$. Since $-{\rm Im}\Pi^{\rm p}$ is a smooth function, $N'$ needs 
not be infinite.

A generalized Laguerre polynomial takes the approximate form for a large $j$,
$L_j^{(\alpha)}(v)\approx j^{\alpha/2}v^{-\alpha/2}e^{v/2}J_\alpha(2\sqrt{jv})$ \cite{BBC},
up to corrections of $1/\sqrt{j}$, $J_\alpha$ being a Bessel function of the first kind. 
Equation~(\ref{d0}) becomes
\begin{eqnarray}
\Delta \rho(m)
\approx\sum_{j=N}^{N'} a_{j}\sqrt{\frac{j(m^2-4m_b^2)}{\Lambda}}^{\alpha} e^{-(m^2-4m_b^2)/(2\Lambda)}
J_\alpha\left(2\sqrt{\frac{j(m^2-4m_b^2)}{\Lambda}}\right),
\label{d1}
\end{eqnarray}
where the variable $v$ has been written as $(m^2-4m_b^2)/\Lambda$ explicitly. Defining the scaling 
variable $\omega\equiv\sqrt{N/\Lambda}$, we have the approximation 
$N'/\Lambda=\omega^2[1+(N'-N)/N]\approx \omega^2$ for a finite $N'-N$. Equation~(\ref{d1}) then 
reduces to
\begin{eqnarray}
\Delta \rho(m)\approx
y\left(\omega \sqrt{m^2-4m_b^2}\right)^{\alpha} J_\alpha\left(2\omega \sqrt{m^2-4m_b^2}\right),
\label{d2}
\end{eqnarray}
as the common Bessel functions 
$J_\alpha(2\sqrt{j(m^2-4m_b^2)/\Lambda})\approx J_\alpha(2\omega \sqrt{m^2-4m_b^2})$ for 
$j=N,N+1,\cdots,N'$ are factored out, and the sum of the unknown coefficients 
$\sum_{j=N}^{N'} a_{j}$ is denoted by $y$. The exponential suppression factor 
$e^{-(m^2-4m_b^2)/(2\Lambda)}=e^{-\omega^2 (m^2-4m_b^2)/(2N)}$ has been replaced by unity for a 
large $N$ in the region with finite $m$ and $\omega$, which we are interested in. 

We stress that a solution to the dispersion relation should be insensitive to the variation 
of the arbitrary scale $\Lambda$, i.e., of $\omega$, which is introduced via the artificial 
variable changes. To realize this insensitivity, we make a Taylor expansion of $\Delta \rho(m_S)$,
\begin{eqnarray}
\Delta \rho(m_S)=\Delta \rho(m_S)|_{\omega=\bar\omega}+
\frac{d\Delta \rho(m_S)}{d\omega}\Big|_{\omega=\bar\omega}(\omega-\bar\omega)+
\frac{1}{2}\frac{d^2\Delta \rho(m_S)}{d\omega^2}\Big|_{\omega=\bar\omega}
(\omega-\bar\omega)^2+\cdots,\label{ta}
\end{eqnarray}
where the constant $\bar\omega$, together with the index $\alpha$ and the coefficient $y$, are 
fixed via the fit of the first term $\Delta \rho(m_S)|_{\omega=\bar\omega}$ to 
$-m_S{\rm Im}\Pi^{\rm p}(m_S)$ in the interval $(2m_b,2m_B)$ of $m_S$. The 
insensitivity to the scaling variable $\omega$ requires the vanishing of the first derivative in 
Eq.~(\ref{ta}), $d\Delta \rho(m_S)/d\omega|_{\omega=\bar\omega}=0$, from which roots of $m_S$ are 
solved. Furthermore, the second derivative $d^2\Delta \rho(m_S)/d\omega^2|_{\omega=\bar\omega}$ 
should be minimal to maximize the stability window around $\bar\omega$, in which $\Delta\rho(m_S)$ 
remains almost independent of $\omega$. It will be seen that only when $m_S$ takes a value close to
the physical Higgs boson mass, can the above requirements be satisfied. Equation~(\ref{d2}) with 
this specific $m_S$ establishes the solution to the dispersion relation in Eq.~(\ref{ge}) with the 
initial condition in the interval $(2m_b,2m_B)$. Once the solution is found, we increase 
the degree $N$ for the polynomial expansion in Eq.~(\ref{d0}) and the scale $\Lambda$
arbitrarily by keeping $\omega=\sqrt{N/\Lambda}$ within the stability window. 
Then all the deductions based on the large $N$ scenario are justified.

\begin{figure}
\begin{center}
\includegraphics[scale=0.25]{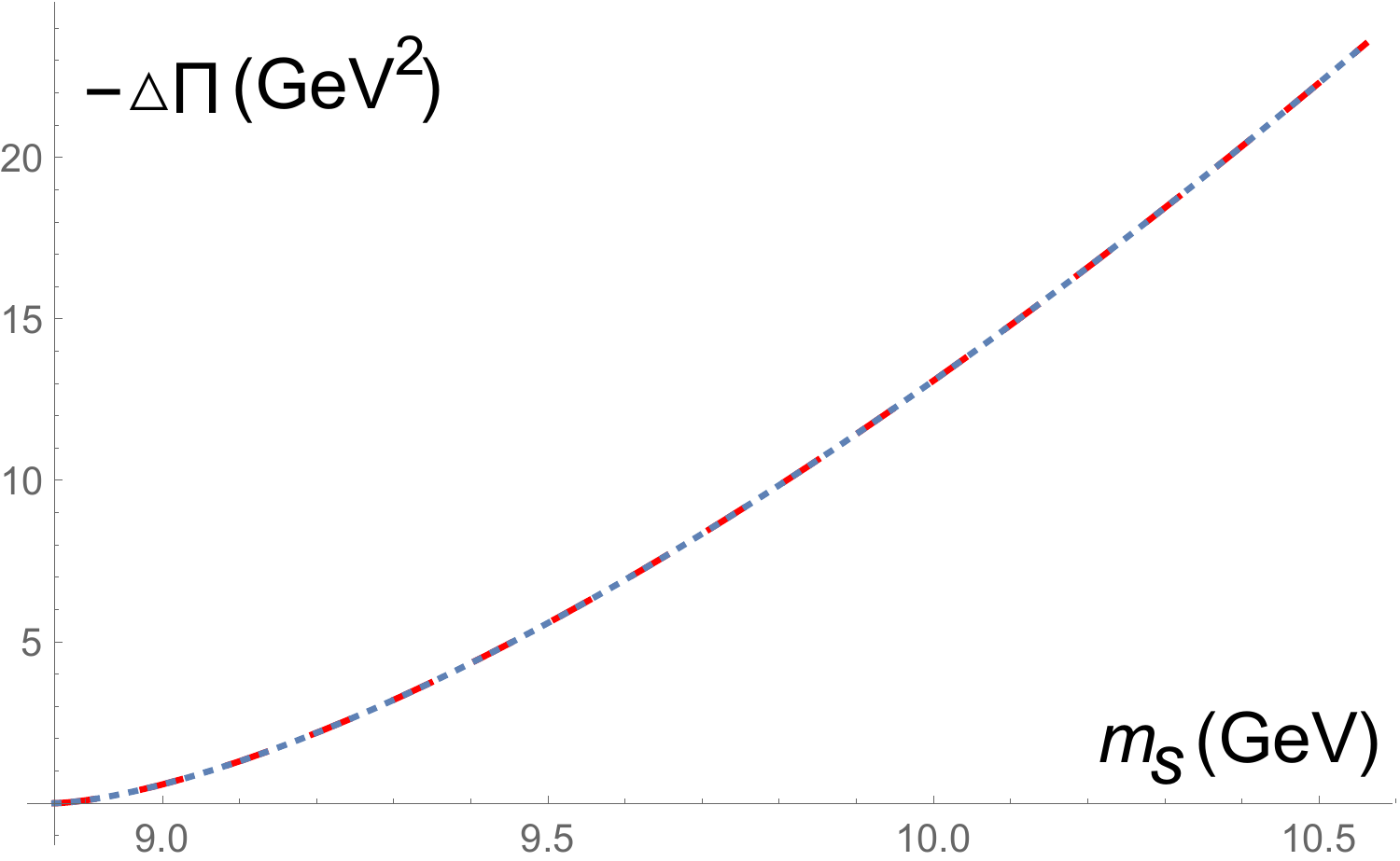}\hspace{1.0cm}
\includegraphics[scale=0.25]{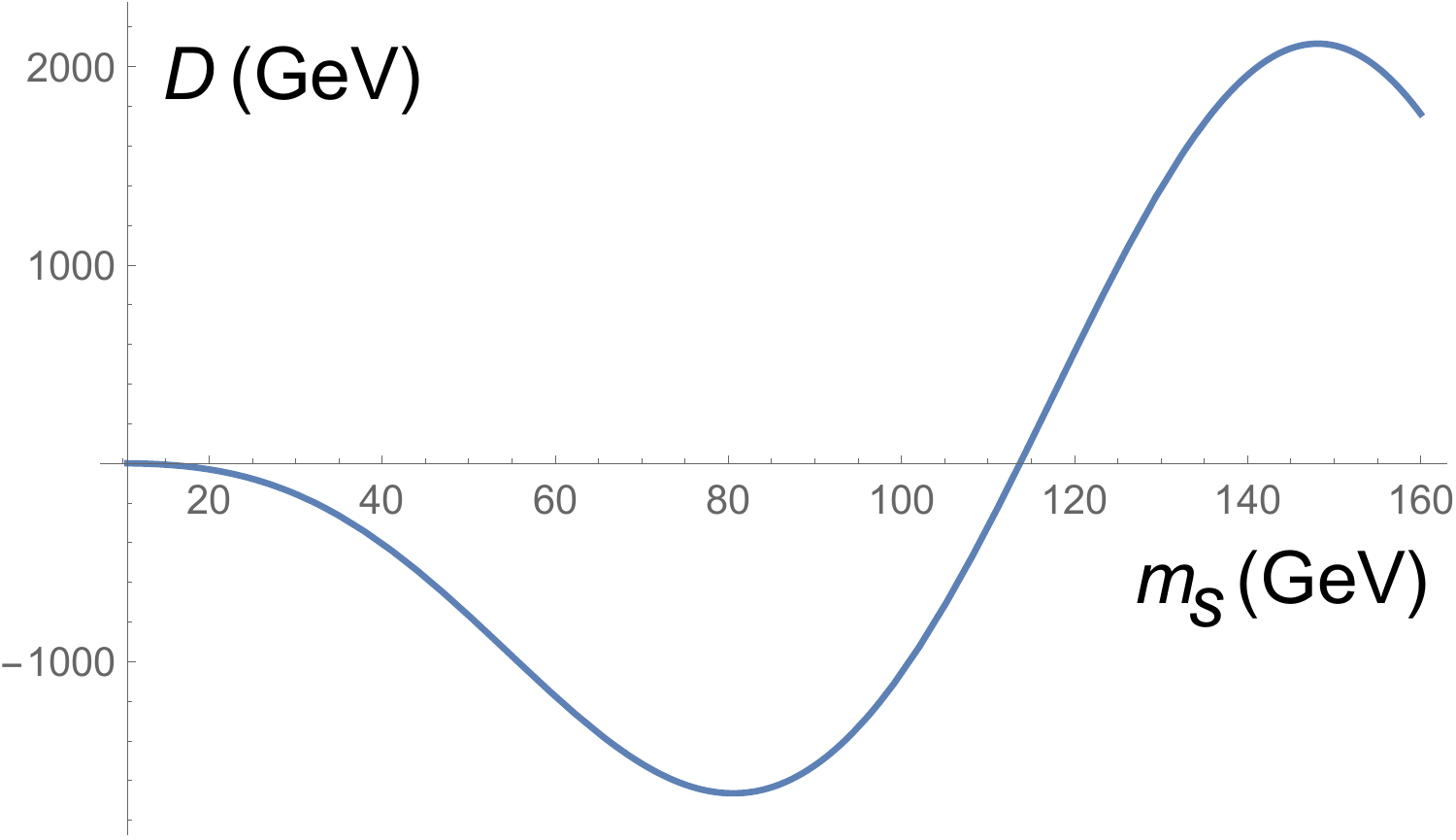}

(a) \hspace{7.0 cm} (b)
\caption{\label{fig2}
(a) Comparison of $-\Delta\Pi(m_S)$ from the fit (dotted line) with the input
${\rm Im}\Pi^{\rm p}(m_S)$ (dashed line) in Eq.~(\ref{p1}) in the interval 
$(2m_b,2m_B)$ of $m_S$. (b) Dependence of the derivative $D(m_S)$ 
in Eq.~(\ref{der}) on $m_S$.}
\end{center}
\end{figure}

Comparing $\Delta \rho(m_S)$ in Eq.~(\ref{d2}) in the limit $m_S\to 2m_b$ with the input 
\begin{eqnarray}
-m_S{\rm Im}\Pi^{\rm p}(m_S)\propto (m_S^2-4m_b^2)^{3/2},\label{p2}
\end{eqnarray}
we read off the index $\alpha=3/2$. It is clear now why we considered the contour integration
of $m^2\Pi(m)$: the corresponding input is proportional to a 
simple power of $m^2-4m_b^2$, so that the index $\alpha$ can be determined unambiguously.
The boundary condition $\Delta\rho(m_S)$ at $m_S=2m_B$ sets the coefficient
\begin{eqnarray}
y=-2m_B{\rm Im}\Pi^{\rm p}(2m_B)
\left[\left(2\omega \sqrt{m_B^2-m_b^2}\right)^{3/2} 
J_{3/2}\left(4\omega \sqrt{m_B^2-m_b^2}\right)\right]^{-1}.\label{dc2}
\end{eqnarray}
The running coupling constant is given by
\begin{eqnarray}
\alpha_s(\mu)=\frac{4\pi}{\beta_0\ln(\mu^2/\Lambda_{\rm QCD}^2)}
\left[1 - \frac{\beta_1\ln\ln(\mu^2/\Lambda_{\rm QCD}^2)}
{\beta_0^2\ln(\mu^2/\Lambda_{\rm QCD}^2)}\right],
\end{eqnarray}
with the coefficients $\beta_0 = 11 - 2n_f/3$ and $\beta_1=2(51-19n_f/3)$. We take the QCD scale 
$\Lambda_{\rm QCD}=0.2$ GeV for the number of active quark flavors $n_f=5$ \cite{Zhong:2021epq}, 
and choose the renormalization scale $\mu=m_S$. Adopting the $b$ quark mass $m_b=4.43$ and the $B$ 
meson mass $m_B=5.28$ \cite{PDG}, we derive $\bar\omega=0.0254$ GeV$^{-1}$ from the best fit of 
Eq.~(\ref{d2}) to $-m_S{\rm Im}\Pi^{\rm p}(m_S)$ from Eq.~(\ref{p1}) in the interval 
$(2m_b,2m_B)$ of $m_S$. The fit result in terms of $-\Delta\Pi(m_S)$ is contrasted with
${\rm Im}\Pi^{\rm p}(m_S)$ in Fig.~\ref{fig2}(a). Their perfect match confirms that the approximate
solution in Eq.~(\ref{d2}) works well, and that other methods for acquiring $\bar\omega$ should 
return similar values: for example, equating $\Delta\Pi(m_S)$ and $-{\rm Im}\Pi^{\rm p}(m_S)$ at 
$m_S=m_B+m_b$ yields $\bar\omega=0.0256$ GeV$^{-1}$, almost identical to 0.0254 GeV$^{-1}$
from the best fit.

As postulated before, the derivative $d\Delta\rho(m_S)/d\omega$ vanishes at $\omega=\bar\omega$, 
i.e.,
\begin{eqnarray}
D(m_S)\equiv \frac{d}{d\omega}\frac{J_{3/2}\left(2\omega \sqrt{m_S^2-4m_b^2}\right)}
{J_{3/2}\left(4\omega \sqrt{m_B^2-m_b^2}\right)}\Big|_{\omega=\bar\omega}=0,\label{der}
\end{eqnarray}
where the factors independent of $\omega$ in the solution have been dropped for simplicity.
Figure~\ref{fig2}(b) shows the dependence of the derivative $D(m_S)$ on $m_S$, which 
reveals an oscillatory behavior. The first root located at $m_S=2m_B=10.56$ GeV, attributed to 
the boundary condition of $\Delta\rho(m_S)$ at this $m_S$, is trivial and bears no physical 
significance. It has been checked that the second derivatives are larger at higher roots, as 
observed in \cite{Li:2023dqi}, so a smaller root is preferred. Therefore, we identify the second 
root at $m_S=114$ GeV as the physical solution of the Higgs boson mass, which deviates from the 
measured one $m_H=(125.25\pm 0.17)$ GeV by only 9\%. The value of $\Delta \Pi(m_S=114\;{\rm GeV})$, 
amounting to about 10\% of the perturbative input ${\rm Im}\Pi^{\rm p}(m_S=114\;{\rm GeV})$, 
indicates a minor nonperturbative contribution to Higgs decays associated with the hadronic 
threshold $2m_B$.

We estimate the theoretical uncertainties involved in the above analysis. The 
variation of the $b$ quark mass in its well-accepted range causes a minor effect: a higher (lower) 
$m_b=4.8$ (4.0) GeV yields the Higgs boson mass 116 (111) GeV. That is, an 8\% change of $m_b$ 
makes less than 3\% impact on the outcome. We mention that the NLO QCD correction to the 
perturbative input is crucial for attaining a sensible Higgs boson mass. 
Without the $O(\alpha_s)$ piece in Eq.~(\ref{p1}), $\bar\omega$ would be as small as 
$1.9\times 10^{-4}$ GeV$^{-1}$, and the Higgs boson mass becomes as large as $1.5\times 10^4$ GeV. 
It is then necessary to examine the dependence on the renormalization scale $\mu$: the choice 
$\mu=m_S/2$ ($\mu=2m_S$) leads to the Higgs boson mass 126 (112) GeV. It hints that
higher-order corrections are under control, and their inclusion into the perturbative input is 
likely to account for the measured $m_H$. This $O(10\%)$ QCD effect from the initial
condition at the $m_b$ scale is quite reasonable.

\section{$Z$ BOSON MASS}

We extract the $Z$ boson mass in the same formalism. A $Z$ boson decays into a $b\bar b$ quark 
pair through the vertex $\gamma_\mu(v_b+a_b\gamma_5)$, where the vector coupling $v_b$ and the 
axial-vector coupling $a_b$ can vary independently in a mathematical point of view. It is noticed 
in perturbative calculations \cite{Schwinger:1973rv,Jersak:1979uv,Chang:1981qq,Gusken:1985nf}
that the axial-vector contribution is less than 10\% of the vector one in the threshold region 
$(2m_b,2m_B)$, and, in particular, completely negligible near the quark-level threshold $2m_b$. 
Hence, we work on the $v_b$ term, namely, the two-point correlation function
\begin{eqnarray}
\Pi_{\mu\nu}(q^2)=i\int d^4xe^{iq\cdot x}
\langle 0|T[J_\mu(x)J_\nu(0)]|0\rangle=(q_\mu q_\nu-g_{\mu\nu}q^2)\Pi(q^2),\label{cuz}
\end{eqnarray}
with the momentum $q$ being injected into the $b$-quark vector current $J_\mu=\bar b\gamma_\mu b$.
Similarly, we keep only the leading term in the OPE for the correlation function $\Pi(q^2)$, i.e., 
the perturbative contribution \cite{Schwinger:1973rv,Jersak:1979uv,Chang:1981qq,Gusken:1985nf},
which gives the imaginary piece
\begin{eqnarray}
{\rm Im}\Pi^{\rm p}(m_V)\propto m_V^2\beta(m_V)[3-\beta^2(m_V)]
\left\{1+\frac{4\alpha_s(m_V)}{3}
\left[\frac{\pi}{2\beta(m_V)}-\frac{3+\beta(m_V)}{4}
\left(\frac{\pi}{2}-\frac{3}{4\pi}\right)\right]\right\},
\label{pz2}
\end{eqnarray}
with the invariant mass $m_V=\sqrt{q^2}$ and the factor $\beta(m_V)=\sqrt{1-4m_b^2/m_V^2}$. We 
have also suppressed the constant prefactor in the above expression, and taken into account
the NLO QCD correction solely, where the renormalization scale $\mu$ in $\alpha_s$ is set to $m_V$. 

The perturbative input in Eq.~(\ref{pz2}) consists of both even and odd functions in $m_V$, 
${\rm Im}\Pi^{\rm p}(m_V)={\rm Im}\Pi_e^{\rm p}(m_V)+{\rm Im}\Pi_o^{\rm p}(m_V)$, so
the unknown function is decomposed into ${\rm Im}\Pi(m_V)={\rm Im}\Pi_e(m_V)+{\rm Im}\Pi_o(m_V)$ 
accordingly. We start with the contour integration of $\Pi(m_V)$ in this case, deriving
\begin{eqnarray}
\int_{4m_b^2}^\infty\frac{\Delta\Pi_e(m)}{m_V^2-m^2}dm^2=0,\;\;\;\;
\int_{4m_b^2}^\infty\frac{\Delta\Pi_o(m)}{m(m_V^2-m^{2})}dm^{2}=0,\label{ze}
\end{eqnarray}
for the even and odd pieces, respectively. The subtracted unknown functions are defined
by $\Delta\Pi_{e,o}(m)={\rm Im}\Pi_{e,o}(m)-{\rm Im}\Pi_{e,o}^{\rm p}(m)$, which are fixed to
$-{\rm Im}\Pi_{e,o}^{\rm p}(m)$ in the interval $(2m_b,2m_B)$ of $m$.
The solutions to Eq.~(\ref{ze}) are constructed in a similar way. The even piece is written as 
\begin{eqnarray}
\Delta \Pi_e(m_V)\approx
y_e\left(\omega \sqrt{m_V^2-4m_b^2}\right)^{\alpha} J_\alpha\left(2\omega \sqrt{m_V^2-4m_b^2}\right),
\label{z2}
\end{eqnarray}
whose comparison with
\begin{eqnarray}
-{\rm Im}\Pi_e^{\rm p}(m_V)\propto 8\pi\alpha_s(2m_b)m_b^2={\rm const.},\label{ei}
\end{eqnarray}
in the limit $m_V\to 2m_b$ specifies the index $\alpha=0$. The odd piece reads
\begin{eqnarray}
\frac{\Delta \Pi_o(m_V)}{m_V}\approx
y_o\left(\omega \sqrt{m_V^2-4m_b^2}\right)^{\alpha} 
J_\alpha\left(2\omega \sqrt{m_V^2-4m_b^2}\right),\label{o2}
\end{eqnarray}
whose comparison with 
\begin{eqnarray}
-\frac{{\rm Im}\Pi_o^{\rm p}(m_V)}{m_V}\propto(m_V^2-4m_b^2)^{1/2},
\end{eqnarray}
in the limit $m_V\to 2m_b$ places the index $\alpha=1/2$. 

\begin{figure}
\begin{center}
\includegraphics[scale=0.25]{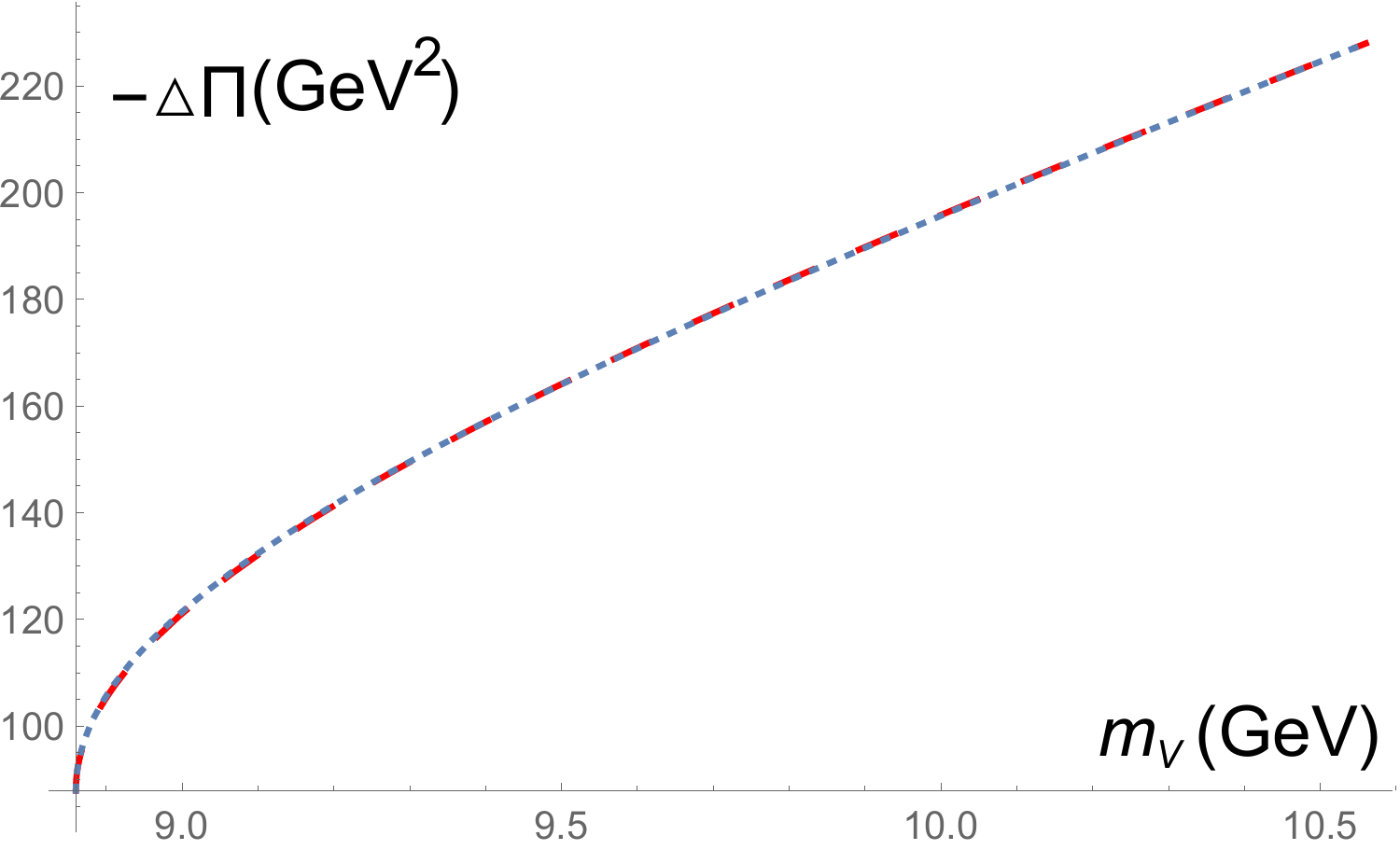}\hspace{1.0 cm} 
\includegraphics[scale=0.25]{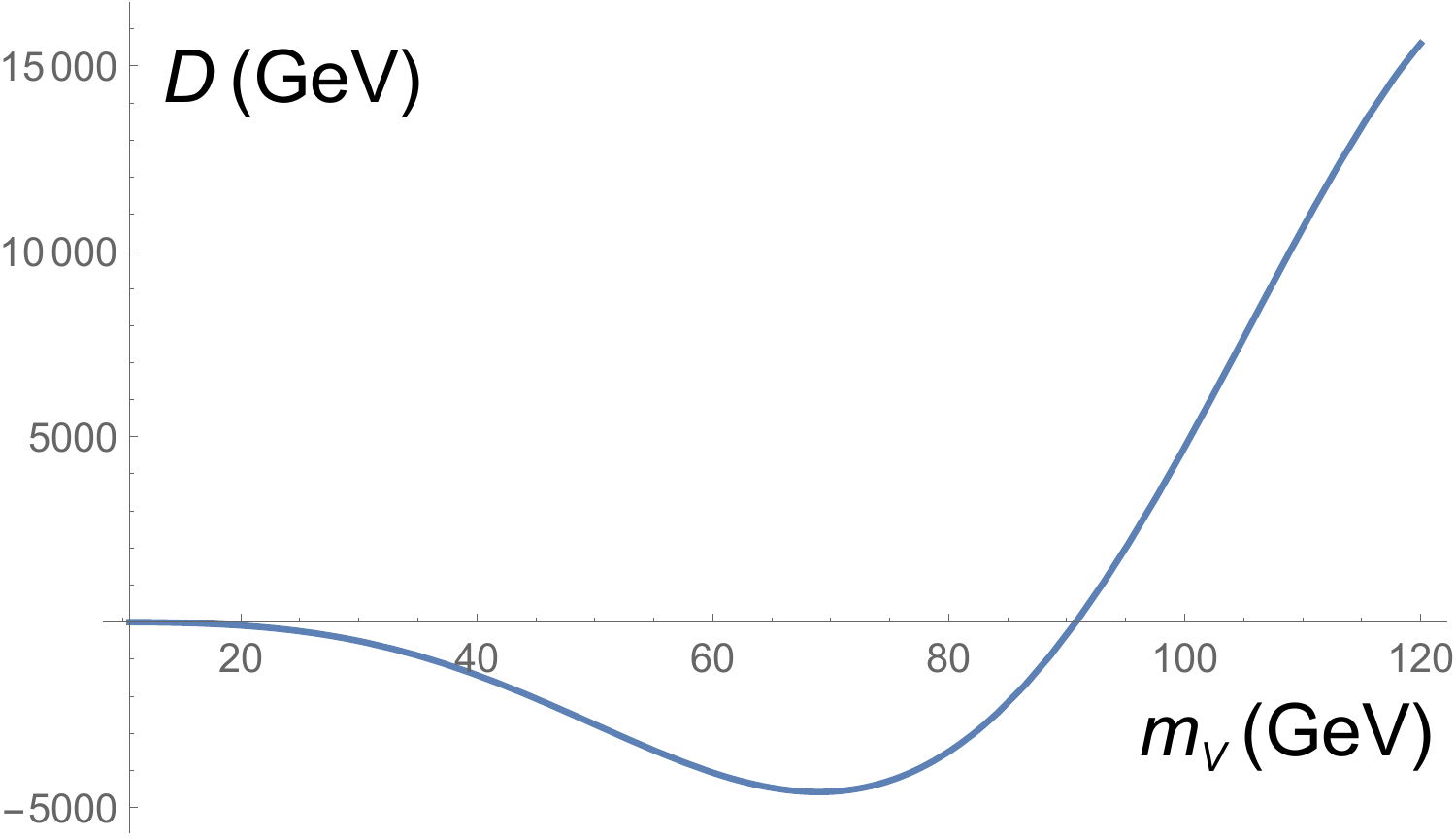}

(a) \hspace{7.0 cm} (b)
\caption{\label{fig3}
(a) Comparison of $-\Delta\Pi(m_V)$ from the fit (dotted line) with the input
${\rm Im}\Pi^{\rm p}(m_V)$ (dashed line) in the interval 
$(2m_b,2m_B)$ of $m_V$. (b) Dependence of the derivative $D(m_V)$ 
in Eq.~(\ref{dez}) on $m_V$.}
\end{center}
\end{figure}

The final solution takes the form
\begin{eqnarray}
\Delta \Pi(m_V)&=&\Delta \Pi_e(m_V)+\Delta \Pi_o(m_V)\nonumber\\
&\approx& y_eJ_0\left(2\omega \sqrt{m_V^2-4m_b^2}\right)+y_om_V
\left(\omega \sqrt{m_V^2-4m_b^2}\right)^{1/2} J_{1/2}\left(2\omega \sqrt{m_V^2-4m_b^2}\right).
\label{zo2}
\end{eqnarray}
The second term in Eq.~(\ref{zo2}) diminishes in the limit $m_V\to 2m_b$, so
the coefficient $y_e$ is related to the initial value $-{\rm Im}\Pi^{\rm p}(m_V=2m_b)$,
\begin{eqnarray}
y_e=-\frac{{\rm Im}\Pi^{\rm p}(2m_b)}{J_0(0)}=-{\rm Im}\Pi^{\rm p}(2m_b).
\end{eqnarray}
The coefficient $y_o$ is obtained from the value $-{\rm Im}\Pi^{\rm p}(m_V=2m_B)$,
\begin{eqnarray}
y_o=-\frac{\left[{\rm Im}\Pi^{\rm p}(2m_B)+y_e J_0\left(4\omega \sqrt{m_B^2-m_b^2}\right)\right]}
{2m_B\left(2\omega \sqrt{m_B^2-m_b^2}\right)^{1/2} J_{1/2}\left(4\omega \sqrt{m_B^2-m_b^2}\right)}.\label{oy2}
\end{eqnarray}
With the same masses $m_b=4.43$ GeV and $m_B=5.28$, we get $\bar\omega=0.0249$ GeV$^{-1}$ 
from the best fit of Eq.~(\ref{zo2}) to $-{\rm Im}\Pi^{\rm p}(m_V)$ in the interval $(2m_b,2m_B)$. 
The fit result $-\Delta\Pi(m_V)$ is contrasted with ${\rm Im}\Pi^{\rm p}(m_V)$ in 
Eq.~(\ref{pz2}) in Fig.~\ref{fig3}(a), where the exact overlap of the two curves 
justifies the approximate solution in Eq.~(\ref{zo2}).

The stability of the solution under the variation of $\omega$ demands the vanishing of the 
derivative 
\begin{eqnarray}
D(m_V)\equiv \frac{1}{y_e}\frac{d\Delta\Pi(m_V)}{d\omega}\Big|_{\omega=\bar\omega}=0,\label{dez}
\end{eqnarray}
where the additional constant $1/y_e$, rendering the differentiated function dimensionless, does 
not affect the locations of roots in $m_V$. Figure~\ref{fig3}(b) displays the dependence of the 
above derivative on $m_V$, where the first root at $m_V=2m_B$ has no physical significance. The 
second root at $m_V=90.8$ GeV corresponds to the physical solution of the $Z$ boson mass, 
which deviates from the measured one $m_Z=(91.1876\pm 0.0021)$ GeV \cite{PDG} by only 0.4\%. 
This result is not sensitive to the variation of the $b$ quark mass, analogous to the Higgs boson 
case. It is observed that the nonperturbative contribution arising from the difference between 
the thresholds $2m_b$ and $2m_B$ also amounts to about 20\% of the perturbative contribution to 
${\rm Im}\Pi(m_Z)$, i.e., the $Z$ boson decay width. We remark that $\Upsilon(4S)$ with the quantum 
numbers $J^{PC}=1^{--}$, the mass 10.58 GeV and the width 20.5 MeV can decay into the $B\bar B$ 
state. To probe such a fine resonance structure in our formalism, more precise OPE for the 
correlation function including QCD condensates \cite{Narison:1994ke} is needed.

Note that the vector coupling $v_b= -1 +4\sin^2\theta_W/3$, with the Weinberg angle $\theta_W$ 
and $\sin\theta_W=g'/\sqrt{g^2+g^{\prime 2}}$, is a constant, given the U(1) coupling $g'$
and the SU(2) coupling $g$ in the SM. The relation $\sin^2\theta_W=1-m_W^2/m_Z^2$ then implies 
that the $W$ and $Z$ boson masses are proportionate. It is interesting to verify 
this proportionality by introducing the $m_V$-dependent coupling
\begin{eqnarray}
v_b(m_V)= -1 +\frac{4}{3}\left(1-\frac{m_W^2}{m_V^2}\right),\label{vb}
\end{eqnarray}
for the fixed $W$ boson mass $m_W=80.377$ GeV into the perturbative input in Eq.~(\ref{pz2}). The 
additional factor $1/m_V^2$ in Eq.~(\ref{vb}) modifies the derivation for the odd piece of the 
solution, which follows the contour integration of $m_V^2\Pi(m_V)/(m_V^2-4m_b^2)$ as in the Higgs 
boson case. The comparison of the solution with $-m_V{\rm Im}\Pi_o^{\rm p}(m_V)/(m_V^2-4m_b^2)$ 
in the interval $(2m_b,2m_B)$ sets the index $\alpha=-1/2$. The 
straightforward steps yield $m_Z=38$ GeV, inconsistent with the measured value. That is, the 
solution $m_V=90.8$ GeV exists, only when the $W$ and $Z$ boson masses are correlated to each 
other, conforming to the Higgs mechanism of the electroweak symmetry breaking. It also means that 
the $W$ boson mass is known, once the $Z$ boson mass is extracted from 
our dispersion analysis and the electroweak couplings are designated.


\section{TOP QUARK MASS}

At last, we come to the determination of the top quark mass. Because a top quark does not 
form a bound state, the correlation functions defined in terms of hadronic matrix elements, such as
the heavy meson lifetimes considered in \cite{Li:2023dqi} for constraining heavy quark masses, 
may not be suitable. Therefore, we extend the formalism developed for neutral meson mixing in
\cite{Li:2022jxc} to the $Q\bar u$-$\bar Qu$ mixing through the box diagrams with a fictitious 
heavy quark $Q$ of an arbitrary mass $m_Q$. The Cabibbo-Kobayashi-Maskawa (CKM) factors associated 
with the intermediate $b\bar b$, $b\bar s$, $s\bar s$, $\cdots$ channels can vary independently in a 
mathematical viewpoint, so we focus on the dispersion relation for the $b\bar b$ channel below. All 
our dispersive studies on the electroweak-scale masses then depend on a single 
quark-level threshold $2m_b$. The box diagrams generate two effective four-fermion operators with 
the $(V-A)(V-A)$ and $(S-P)(S-P)$ structures. It suffices to pick up the contribution from the 
former, whose imaginary piece in perturbative evaluations is expressed as \cite{Cheng,BSS}
\begin{eqnarray}
{\rm Im}\Pi^{\rm p}(m_Q)&\propto&C_2(m_Q)
\frac{\sqrt{m_Q^2-4m_b^2}}{m_Q(m_W^2-m_b^2)^2}
\left[2\left(1+\frac{m_b^4}{4m_W^4}\right)
m_W^2(m_Q^2-4m_b^2)-6m_b^2(m_Q^2-2m_b^2)\right].
\label{bij}
\end{eqnarray} 
We have kept only the Wilson coefficient $C_2(\mu)$, which dominates over $C_1(\mu)$ at the relevant 
renormalization scale $\mu=m_Q$.

\begin{figure}
\begin{center}
\includegraphics[scale=0.25]{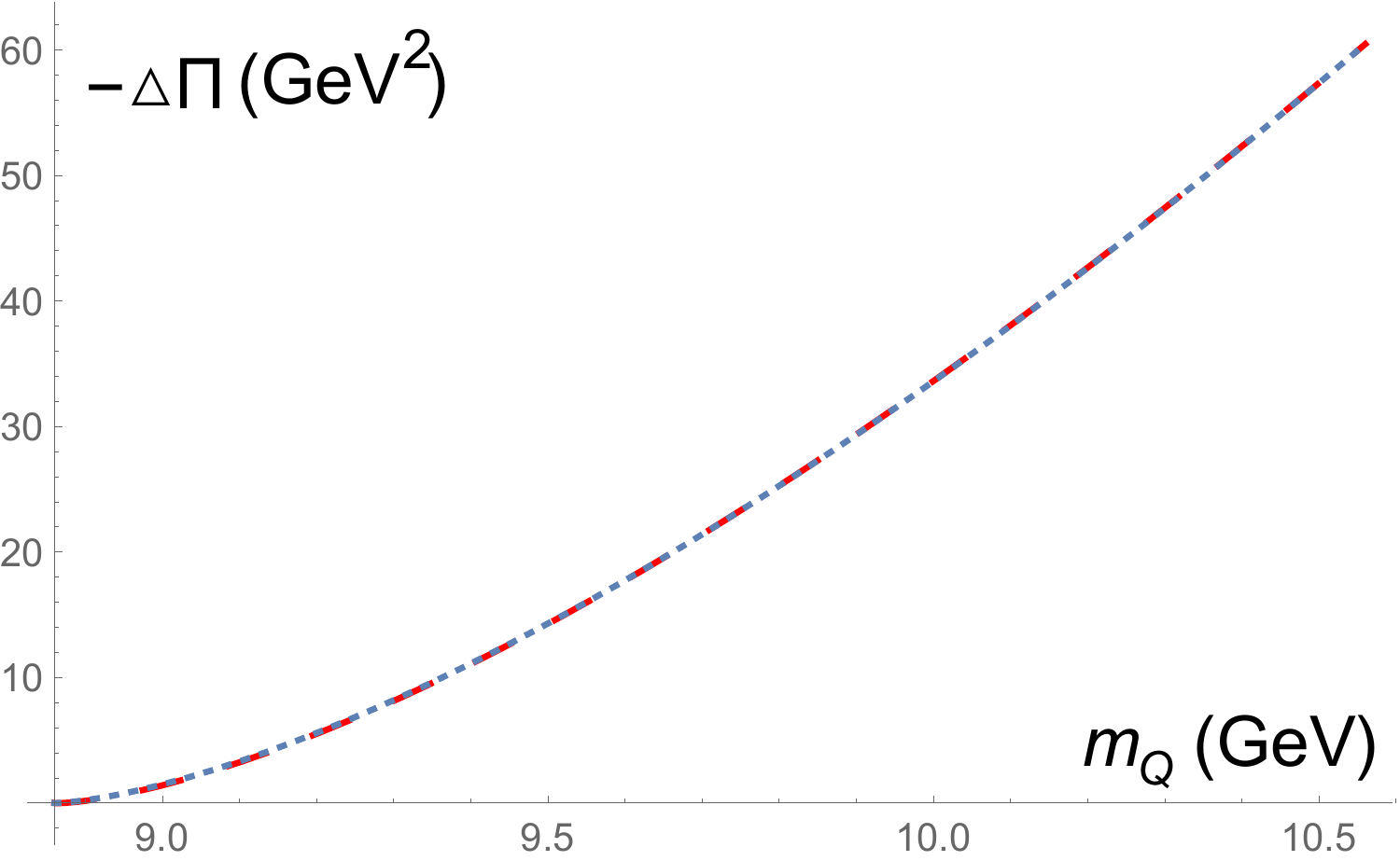}\hspace{1.0 cm} 
\includegraphics[scale=0.25]{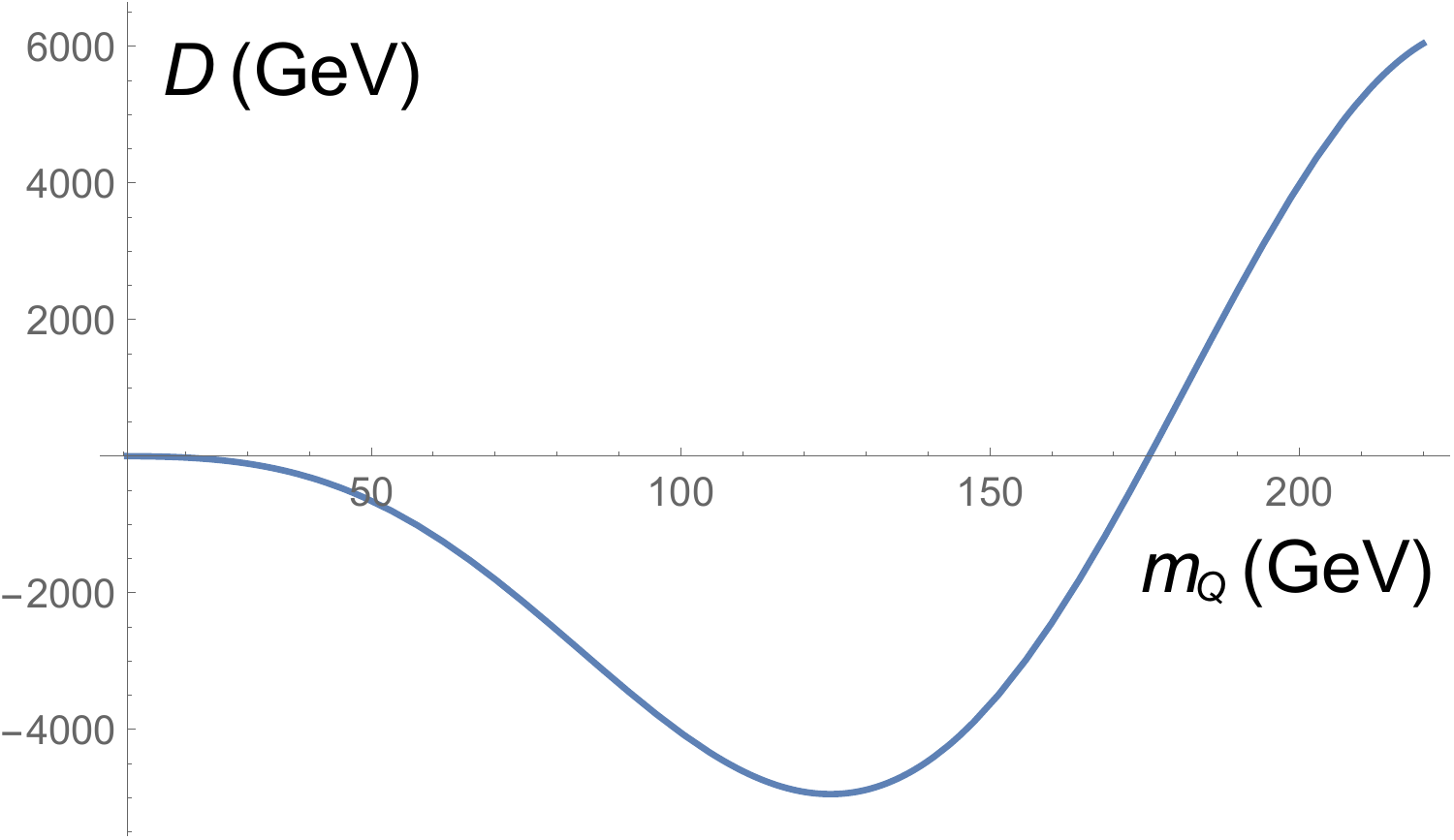}

(a) \hspace{7.0 cm} (b)
\caption{\label{fig4} 
(a) Comparison of $-\Delta\Pi(m_Q)$ from the fit (dotted line) with the input
${\rm Im}\Pi^{\rm p}(m_Q)$ (dashed line) in the interval 
$(2m_b,2m_B)$ of $m_V$. (b) Dependence of the derivative $D(m_Q)$ 
in Eq.~(\ref{der}) on $m_Q$.}
\end{center}
\end{figure}

The second term in the square brackets of Eq.~(\ref{bij}) is down by a tiny ratio $m_b^2/m_W^2$, 
so the behavior of Eq.~(\ref{bij}) in $m_Q$ is governed by the first term. It suggests an 
analysis similar to that for the Higgs boson, and the solution of the unknown subtracted function 
$\Delta\rho(m_Q)$ is the same as Eq.~(\ref{d2}). The comparison of Eq.~(\ref{d2}) with Eq.~(\ref{bij}) 
in the region around $m_Q\sim 2m_b$ then specifies the index $\alpha=3/2$. The coefficient $y$ 
fixed by the boundary condition at $m_Q=2m_B$ is identical to Eq.~(\ref{dc2}). The same 
$b$ quark and $B$ meson masses give $\bar\omega=0.0164$ GeV$^{-1}$ from the best fit of 
Eq.~(\ref{d2}) to the perturbative input $-m_Q{\rm Im}\Pi^{\rm p}(m_Q)$
in the interval $(2m_b,2m_B)$ of $m_Q$. We present the fit result in terms of 
$-\Delta\Pi(m_Q)$ and the perturbative input ${\rm Im}\Pi^{\rm p}(m_Q)$ in Fig.~\ref{fig4}(a)
with excellent match between them. The derivative $d\Delta\rho(m_Q)/d\omega$ vanishes at 
$\omega=\bar\omega$ as in Eq.~(\ref{der}). Figure~\ref{fig4}(b) exhibits the dependence of the 
derivative $D(m_Q)$ on $m_Q$, where the second root at $m_Q=176$ GeV 
corresponds to the physical top quark mass, with only 2\% deviation from the measured 
one $m_t=(172.69\pm 0.30)$ GeV \cite{PDG}. 

The solved top quark mass is more sensitive to the variation of $m_b$ compared to the Higgs and 
$Z$ boson cases: choosing a slightly smaller $m_b=4.42$ GeV would lower the root of $m_Q$ to 169 
GeV. Decreasing the renormalization scale in the Wilson coefficient $C_2$ a bit to $\mu=0.98 m_Q$ 
leads to $m_t=173$ GeV in agreement with the data. It hints that the top quark mass can be well accommodated within the theoretical uncertainty of the calculation.
To be cautious, we have repeated the above derivation for the $B_d$ meson mixing via the 
$c\bar c$ channel. The difference is that the HQE contributions from the effective weak Hamiltonian
\cite{Beneke:1998sy,Ciuchini:2003ww}, instead of from the box diagrams, are input.
Taking the $c$ quark mass $m_c=1.35$ GeV \cite{Li:2023dqi} and the $D$ meson mass $m_D=1.86$ GeV \cite{PDG}, we deduce the $b$ quark mass $m_b\approx 4.7$ GeV, close to the values assumed in 
the present work and extracted from the study of heavy meson lifetimes \cite{Li:2023dqi}.  
This result, to be published elsewhere, solidifies the determination of the top quark mass
and the consistency of our approach.




\section{CONCLUSION}

We have demonstrated that dispersive analyses of physical observables involving heavy particles can 
disclose stringent connections between their high-energy and low-energy behaviors. The strategy 
is to treat the dispersion relation obeyed by a correlation function as an inverse problem with 
reliable perturbative inputs in the high-energy region. Two arbitrary parameters were introduced 
into the formalism: the lowest degree $N$ for the Laguerre polynomial expansion and the variable 
$\Lambda$, which scales the heavy particle mass in dispersive integrals. The essential feature is 
that the solution for a physical observable depends on the ratio $\omega^2=N/\Lambda$, which it must 
be insensitive to. This criterion can be met, only when the heavy particle takes a specific 
mass. Once the solution with the physical mass is established, both $N$ and $\Lambda$ can be 
extended to sufficiently high values by keeping $\omega$ in the stability window. All the large $N$ 
approximations assumed in solving the dispersion relation are then justified. We emphasize that no 
a priori information of the considered heavy particle was included: the fictitious scalar mass $m_S$, 
vector mass $m_V$ and heavy quark mass $m_Q$ appearing in the dispersion relations are completely 
arbitrary, and the $b$ quark mass is the only input parameter. Hence, the emergence of the Higgs 
boson, $Z$ boson and top quark masses as the stable solutions is not coincidence, but a necessary
consequence of highly nontrivial constraints imposed by the dispersion relations.

We point out that the strong interaction plays an important role here: the initial 
conditions in the interval bounded by the quark- and hadron-level thresholds are indispensable for 
the existence of nontrivial solutions. Our previous analysis has manifested the correlation 
among the masses of a strange quark, charm quark, bottom quark, muon and $\tau$ lepton through 
the dispersion relations for heavy meson decay widths. Together with the present work, we claim 
that the particle masses from 0.1 GeV up to the electroweak scale may be understood by means of
the internal consistency of SM dynamics, and that at least some of the SM parameters are not free, 
but constrained dynamically. Our observation provides a natural explanation of the flavor 
structures of the SM without resorting to any new symmetries, interactions or mechanism, as attempted 
intensively in the literature. We have taken the perturbative inputs up to NLO in QCD, and 
the current determinations of the particle masses deviate from the measured ones at percent 
level. More precise predictions will be pursued by taking into account subleading contributions 
in inputs, and theoretical uncertainties will be investigated more thoroughly in the future.


\section*{Acknowledgement}

This work was supported in part by National Science and Technology Council of the Republic
of China under Grant No. MOST-110-2112-M-001-026-MY3.


\end{document}